\documentclass[12pt]{article}
\usepackage[dvips]{graphicx}
\usepackage{amsmath,amssymb,amsfonts}

\begin{document}

\title{Effect of an absorbing medium on particle oscillations}
\author{V.I.Nazaruk\\
Institute for Nuclear Research of RAS, 60th October\\
Anniversary Prospect 7a, 117312 Moscow, Russia.*}

\date{}
\maketitle
\bigskip

\begin{abstract}
Particle oscillations in absorbing matter are considered. The approach based on the optical potential is shown to be inapplicable in the strong absorption region. Models with Hermitian Hamiltonian are analyzed. They give an increase of the process width in comparison with the model based on the optical potential.

\end{abstract}

\vspace{5mm}
\vspace{1cm}

*E-mail: nazaruk@inr.ru
\newpage
\setcounter{equation}{0}
\section{Introduction}
In particle oscillations in the medium absorption can play an important role, for example, in the $K^0\bar{K}^0$ [1-4] and $n\bar{n}$ [5-8] oscillations. In this paper we consider $n\bar{n}$ transitions in the medium followed by annihilation 
\begin{equation}
n\rightarrow \bar{n}\rightarrow M.
\end{equation}
Here $M$ are the annihilation mesons. The reason for considering this process is that the absorption (annihilation) of $\bar{n}$ is extremely strong.

In the standard approach (later on referred to as a potential model) the $\bar{n}$-medium interaction is described by antineutron optical potential $U_{\bar{n}}$. We have objections to this model (Sect. 2). In Sect. 3 the alternative models based on the field-theoretical approach are considered. For these models two possibilities exist: a model with bare (Sect. 3.1) and dressed (Sect. 3.3) propagators. (In the latter case we come to the $S$-matrix problem formulation.) In the models with bare and dressed propagators we directly calculate the off-diagonal matrix element without using the optical potential.

The results are compared in Sect. 4. The potential model contains double counting. This has been proved in the standard $S$-matrix approach. This fact should be emphasized particularly.  

In Sect. 5 the results are summarized. The problems of the models based on the $S$-matrix approach are pointed out as well. The restriction on the free-space $n\bar{n}$ oscillation time $\tau $ critically depends on the description of absorption. In this regard, the main goal of this paper is to consider the absorption model itself. 

\section{Potential model}
We consider process (1). In the standard approach [5-7] the $n\bar{n}$ transitions in the medium are described by Schrodinger equations:
\begin{eqnarray}
(i\partial_t-H_0)n(x)=\epsilon _{n\bar{n}}\bar{n}(x),\nonumber\\
(i\partial_t-H_0-V)\bar{n}(x)=\epsilon _{n\bar{n}}n(x),\nonumber\\
H_0=-\nabla^2/2m+U_n,\nonumber\\
V=U_{\bar{n}}-U_n={\rm Re}U_{\bar{n}}+i{\rm Im}U_{\bar{n}}-U_n,
\end{eqnarray}
${\rm Im}U_{\bar{n}}=-\Gamma /2$, $\bar{n}(0,{\bf x})=0$. Here $U_n$ and $U_{\bar{n}}$ are the potential of $n$ and the optical potential of $\bar{n}$, respectively; $\epsilon _{n\bar{n}}$ is a small parameter with $\epsilon _{n\bar{n}}=1/\tau $, where $\tau $ is the free-space $n\bar{n}$ oscillation time, $\Gamma $ being the annihilation width of $\bar{n}$.

In the lowest order in $\epsilon _{n\bar{n}}$ the process width is [5-7]
\begin{equation}
\Gamma _{pot}=\epsilon _{n\bar{n}}^2\frac{1}{({\rm Re}V)^2+(\Gamma /2)^2}\Gamma .
\end{equation}
$U_{\bar{n}}$ is the basic element of the model. In this connection the following problems arise:

1. The optical model was developed for the Schrodinger type equations. The physical meaning of ${\rm Im}U_{\bar{n}}$ follows from the corresponding continuity equation. Coupled Eqs. (2) give rise to the following equation:
\begin{equation}
(\partial_t^2+i\partial_t(V+2H_0)-H_0^2-H_0V+ \epsilon _{n\bar{n}}^2)n(x)=0.
\end{equation}
The continuity equation cannot be derived from (4). 

2. To get $\Gamma _{pot}$, the optical theorem or condition of probability conservation are used. However, the $S$-matrix is essentially non-unitary.

3. The structure and $\Gamma $-dependence of (3) provoke some objections. Due to this an alternative model should be considered.

\section{Field-theoretical approach}
The interaction Hamiltonian of process (1) is given by
\begin{eqnarray}
{\cal H}_I={\cal H}_{n\bar{n}}+{\cal H},\nonumber\\
{\cal H}_{n\bar{n}}=\epsilon _{n\bar{n}}\bar{\Psi }_{\bar{n}}\Psi _n+H.c.,
\end{eqnarray}
where ${\cal H}_{n\bar{n}}$ and ${\cal H}$ are the Hamiltonians of $n\bar{n}$ conversion and the $\bar{n}$-medium interaction, respectively. The background neutron potential is included in the neutron wave function:
\begin{equation}
n(x)=\Omega ^{-1/2}\exp (-ipx), 
\end{equation}
$p=(\epsilon ,{\bf p})$, $\epsilon ={\bf p}^2/2m+U_n$.

\subsection{Model with a bare propagator}
The $n\bar{n}$ conversion comes from the exchange of Higs bosons with $m_H>10^5$ GeV. The $\bar{n}$ annihilates in a time $\tau _a\sim 1/\Gamma $. We deal with a two-step process with a characteristic time $\tau _a$.

The general definition of the antineutron annihilation amplitude $M_a$ is given by
\begin{equation}
<\!M0\!\mid T\exp (-i\int dx{\cal H}(x))-1\mid\!0\bar{n}_{p}\!>=
N(2\pi )^4\delta ^4(p_f-p_i)M_a.
\end{equation}
Here $\mid\!0\bar{n}_{p}\!>$ is the state of the medium containing the $\bar{n}$ with the 4-momentum $p=(\epsilon ,{\bf p})$; $<\!M\!\mid $ denotes the annihilation mesons, $N$ includes the normalization factors of the wave functions. The antineutron annihilation width $\Gamma $ is expressed 
through $M_a$:
\begin{equation}
\Gamma =N_1\int d\Phi \mid\!M_a\!\mid ^2,
\end{equation}
where $N_1$ is the normalization factor.

The amplitude of process (1) $M_1$ is given by
\begin{equation}
<\!M0\!\mid T\exp (-i\int dx{({\cal H}_{n\bar{n}}(x)+\cal H}(x)))-1\mid\!0n_{p}\!>=
N(2\pi )^4\delta ^4(p_f-p_i)M_1.
\end{equation}

\begin{figure}[h]
  {\includegraphics[height=.25\textheight]{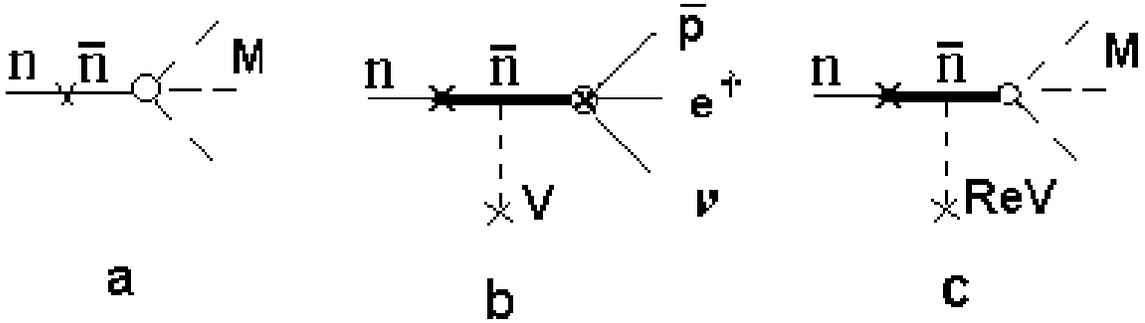}}
  \caption{{\bf a} $n\bar{n}$ transition in the medium followed by annihilation. The 
annihilation is shown by a circle. The propagator is bare {\bf b} $n\bar{n}$ transition in the medium followed by decay
{\bf c} The same as {\bf a} but the antineutron
propagator is dressed (see text)}
\end{figure}

In the lowest order in ${\cal H}_{n\bar{n}}$ for the process amplitude $M_1$ one obtains (see Fig. 1a) 
\begin{equation}
M_1=\epsilon _{n\bar{n}}G_0M_a,
\end{equation}
\begin{equation}
G_0=\frac{1}{\epsilon _{\bar{n}} -{\bf p}_{\bar{n}}^2/2m-U_n+i0},
\end{equation}
where $G_0$ is the antineutron propagator. Since ${\bf p}_{\bar{n}}={\bf p}$, $\epsilon _{\bar{n}}=\epsilon $, then $G_0\sim 1/0$. $M_a$ contains all the $\bar{n}$-medium interactions followed by annihilation including antineutron rescattering in the initial state. So in this case the antineutron propagator is bare. 

We deal with infrared singularity. For solving the problem a field-theoretical approach with a finite time interval has been proposed [9]. The process (1) probability was found to be [10]
\begin{equation}
W(t)\approx W_f(t)=\epsilon _{n\bar{n}}^2t^2,
\end{equation}
where $W_f$ is the free-space $n\bar{n}$ transition probability. Equation (12) leads to a very strong restriction on the free-space $n\bar{n}$ oscillation time: $\tau =10^{16}$ yr.

\subsection{Absorption in the intermediate state}
Starting from (5) and (6) we have drawn the singular amplitude $M_1$. To gain a better understanding of the problem, we consider the $n\bar{n}$ transitions in the medium followed by $\beta ^+$-decay:
\begin{equation}
n\rightarrow \bar{n}\rightarrow \bar{p}e^+\nu .
\end{equation}
The neutron wave function is given by (6). The interaction Hamiltonian is
\begin{equation}
{\cal H}_I={\cal H}_{n\bar{n}}+{\cal H}_W+V\bar{\Psi }_{\bar{n}}\Psi _{\bar{n}},
\end{equation}
where $V$ is defined by (2), ${\cal H}_W$ is the Hamiltonian of the decay $\bar{n}\rightarrow \bar{p}e^+\nu $. The process amplitude is non-singular (see (15) below) and we use the $S$-matrix approach. In the lowest order in ${\cal H}_{n\bar{n}}$ the amplitude $M_2$ (see Fig. 1b) is given by
\begin{eqnarray}
M_2=\epsilon _{n\bar{n}}GM_d,\nonumber\\
G=\frac{1}{\epsilon _{\bar{n}}-{\bf p}_{\bar{n}}^2/2m-U_{\bar{n}}+i0}=\frac{1}{\epsilon -{\bf p}^2/2m-(U_n+V)+i0}=-\frac{1}{V},
\end{eqnarray}
where $M_d$ is the amplitude of the $\beta ^+$-decay, $G$ is the antineutron propagator.

The process width $\Gamma _2$ is
\begin{equation}
\Gamma _2=\frac{\epsilon _{n\bar{n}}^2}{\mid\!V\!\mid ^2}\Gamma _d,
\end{equation}
where $\Gamma _d$ is the width of the $\beta ^+$-decay. The propagator is dressed due to the additional field $V$. There are no questions connected with $U_{\bar{n}}$ since $G$ is the propagator of Schrodinger equation.

\subsection{Model with a dressed propagator}
We return to process (1). Let us try to compose a model with a dressed propagator. By analogy with (14) in the Hamiltonian ${\cal H}$ (see (5)) we separate out the scalar field $V_1$:
\begin{equation}
{\cal H}=V_1\bar{\Psi }_{\bar{n}}\Psi _{\bar{n}}+{\cal H}_a,
\end{equation}
where ${\cal H}_a$ is the annihilation Hamiltonian. Now the antineutron annihilation amplitude $M_{an}$ is defined through ${\cal H}_a$:
\begin{equation}
<\!M0\!\mid T\exp (-i\int dx{\cal H}_a(x))-1\mid\!0\bar{n}_{p}\!>=
N(2\pi )^4\delta ^4(p_f-p_i)M_{an}.
\end{equation}
The interaction Hamiltonian is given by
\begin{equation}
{\cal H}_I={\cal H}_{n\bar{n}}+V_1\bar{\Psi }_{\bar{n}}\Psi _{\bar{n}}+
{\cal H}_a.
\end{equation}               
In the lowest order in ${\cal H}_{n\bar{n}}$ the amplitude of process (1) is 
\begin{eqnarray}
M_{s}=\epsilon _{n\bar{n}}G_dM_{an},\nonumber\\
G_d=G_0+G_0V_1G_0+...=
\frac{1}{(1/G_0)-V_1+i0}=-\frac{1}{V_1}.
\end{eqnarray}
The antineutron propagator $G_d$ is dressed. $V_1$ plays the role of antineutron self-energy $\Sigma $. $M_{s}$ corresponds to the first order in ${\cal H}_{n\bar{n}}$ and all the orders in $V_1$ and ${\cal H}_a$. Compared to (7),  $M_{an}$ is calculated through the reduced Hamiltonian ${\cal H}_a$ instead of ${\cal H}$; otherwise $V_1=0$ and we arrive at the amplitude (10).

The process width $\Gamma _{s}$ is 
\begin{eqnarray}
\Gamma _{s}=N_1\int d\Phi \mid\!M_{s}\!\mid ^2=\frac{\epsilon _{n\bar{n}}^2}{\mid\!V_1\!\mid ^2}\Gamma _{an},\nonumber\\
\Gamma _{an}= N_1\int d\Phi \mid\!M_{an}\!\mid ^2.
\end{eqnarray}

The amplitude $M_{s}$ is non-singular because the propagator is dressed. The antineutron self-energy $\Sigma =V_1$ appears due to separation of the field $V_1$. This procedure seems to be artificial and unjustified as well as definition of the $M_{an}$. There are no similar problems for process (13) since the self-energy and decay of $\bar{n}$ are generated by different fields ${\cal H}_W$ and $V$. This point should be given particular emphasis. In any case $\Gamma _{an}\sim \Gamma $, and so 
\begin{equation}
\Gamma _{s}\sim \Gamma _{an}\sim \Gamma .
\end{equation} 
            
\section{Comparison with potential model}
\subsection{Double counting in the potential model}
First of all we compare the potential model with the model with a dressed propagator. In (21) we have to take the same parameters as in the potential model: $V_1=V$ and $\Gamma _{an}=\Gamma $. Then we get
\begin{equation}
\Gamma _{s}=\epsilon _{n\bar{n}}^2\frac{1}{({\rm Re}V)^2+(\Gamma /2)^2}\Gamma .
\end{equation}
Equation (23) coincides with (3): $\Gamma _{s}=\Gamma _{pot}$. By means of the model with a dressed propagator we have obtained $\Gamma _{pot}$. The antineutron annihilation width $\Gamma $ is involved in the propagator (see (20), where $V_1=V$) as well as vertex function which means double counting. 

The same conclusion has been done in [8]. It was shown that double counting leads to full cancellation of the leading terms.  
However, in [8] the consideration was qualitative and performed on the finite time interval. Equation (23) reproduces (3) exactly.

\subsection{Model with Hermitian Hamiltonian}
As proved earlier, the model with dressed propagator is unjustified. Nevertheless, the correction of the type (17) cannot be excluded. As an alternative to the model with bare propagator we consider the model with dressed propagator (see Fig. 1c). The model is simple: $U_n$ and $U_{\bar{n}}$ are the real potentials of $n$ and $\bar{n}$, respectively; annihilation included in the vertex function only; energy gap ${\rm Re}V$ leads to the process suppression. As with model with bare propagator, the Hamiltonin is Hermitian.        

In (21) we take $V_1={\rm Re}V$ (in this case the Hamiltonin is Hermitian), and $\Gamma _{an}=\Gamma $. The process width $\Gamma _{s}$ is
\begin{equation}
\Gamma _{s}=\frac{\epsilon _{n\bar{n}}^2}{({\rm Re}V)^2}\Gamma .
\end{equation}   
The model described above is the most realistic variant of the model with dressed propagator.

Therefore, $\Gamma _{s}\sim \Gamma $. For the $K^0\bar{K}^0$ transitions in the medium followed by decay and regeneration of the $K^0_{S}$-component an identical $\Gamma $-dependence takes plays [11,12]. In the potential model $\Gamma _{pot}\sim \Gamma $ only at light absorption. Indeed, if $\Gamma /2\ll \mid\!{\rm Re}V\!\mid$, then
\begin{equation}
\Gamma _{pot}=\frac{\epsilon _{n\bar{n}}^2}{({\rm Re}V)^2}\Gamma 
\left[1-\left(\frac{\Gamma }{2{\rm Re}V}\right)^2\right]. 
\end{equation}
In the first approximation (25) coincides with (24). This agreement was expected since the dominant role was played by ${\rm Re}U_{\bar{n}}$.

If $\Gamma /2\gg \mid\!{\rm Re}V\!\mid$,
\begin{equation}
\Gamma _{pot}=\frac{4\epsilon _{n\bar{n}}^2}{\Gamma }.
\end{equation}
$\Gamma _{pot}\sim 1/\Gamma $, whereas $\Gamma _{s}\sim \Gamma $.

The difference in the results is seen from the ratio  
\begin{equation}
r=\frac{\Gamma _{s}}{\Gamma _{pot}}=1+\left(\frac{\Gamma }{2{\rm Re}V}\right)^2.
\end{equation}
If $\mid\!{\rm Re}V\!\mid=\Gamma /2$, then $r=2$. If $\mid\!{\rm Re}V\!\mid=\Gamma /4$, then $r=5$. When $\mid\!{\rm Re}V\!\mid$ decreases, $\Gamma _{s}$ and $r$ increase.

We conclude: (1) The smaller $\mid\!{\rm Re}V\!\mid$ (antineutron self-energy), the greater the difference in the results. It is a maximum for the model with a bare propagator. (2) In the strong absorption region $\Gamma _{pot}\sim 1/\Gamma $, whereas $\Gamma _{s}\sim \Gamma $. (3) The potential model contains double counting.
These conclusions are also true for the model with bare propagator since it is the limiting case $V_1\rightarrow 0$. These conclusions do not depend on the specific models of the blocks $M_a$ and $M_{an}$.  

For the realistic parameters $\Gamma =100$ MeV and $\mid\!{\rm Re}V\!\mid=10$ MeV, the lower limit on the free-space $n\bar{n}$ oscillations time is $\tau =1.2\times 10^{9}\; {\rm s}$. When $V_1=0$, the model with a dressed propagator converts to the model with a bare propagator. It gives $\tau = 10^{16}\; {\rm yr}$. On the basis of this one can accept that the lower limit on the free-space $n\bar{n}$ oscillations time is in the range $10^{16}\; {\rm yr}>\tau >1.2\times 10^{9}\; {\rm s}$.

Finally, in the strong absorption region the model with an optical potential is inapplicable. In our models we calculate directly off-diagonal matrix element. The optical potential is not used. (Note that in the case of Hermitian Hamiltonian the optical theorem is applicable.) 

\section{Conclusion}
The model based on the optical potential compared with direct calculation of off-diagonal matrix element. The potential model is applicable only in the case of slight absorption. 

If absorption is strong, the potential model is inapplicable: (1) It contains double counting. (2) The $\Gamma $-dependence of the result is inverse: $\Gamma _{pot}\sim 1/\Gamma $, whereas $\Gamma _{s}\sim \Gamma $. (3) The physical meaning of ${\rm Im}U_{\bar{n}}$ is uncertain. (4) The using of the optical theorem or condition of probability conservation contradicts the fact that the $S$-matrix is essentially non-unitary.

The field-theoretical approach is free from drawback mentioned above. Two variant of the models have been considered: the model with bare and dressed propagators. (In the latter case we come to the $S$-matrix problem formulation.) If the scalar field $V_1\rightarrow 0$ (the antineutron self-energy $\Sigma \rightarrow 0$), the model with a dressed propagator converts to the model with a bare propagator and so the results are valid for the model with bare propagator as well. In both variants the optical potential is not used. The amplitudes of annihilation $M_a$ and $M_{an}$ are defined through Hermitian Hamiltonians.

The chief drawback in the model with a dressed propagator is that the procedure of separation of $V_1$ (or ${\rm Re}V$) is artificial and unjustified. There are a lot of arguments in favor of the model with a bare propagator [10]. The only objection to this model is that it gives the result which essentially differs from the result of the potential model. The potential model has been considered above. 

In our opinion the model with a bare propagator is preferable. The model with dressed propagator has been considered for the study of process since the problem is of a great nicety. It also gives the conservative limit $\tau =1.2\times 10^{9}\; {\rm s}$. 

In the oscillation of other particles the difference between $\Gamma _{s}$ and $\Gamma _{pot}$ is less, however this difference can be essential for the problem under study. Specifically, for the $K^0_{S}$ regeneration the model with Hermitian Hamiltonian [13] gives the reinforcement as well.

\end{document}